\documentclass{npw}
\usepackage{graphicx}

\newcommand{\bra}{\langle}
\newcommand{\ket}{\rangle}

\newcommand{\beit}{\begin{itemize}}
\newcommand{\enit}{\end{itemize}}

\def\m@thcombine#1#2{%
  \setbox0=\hbox{$#1$}
  \setbox1=\hbox{$#2$}
  \ifdim\wd0>\wd1
    \setbox0=\hbox to\wd1{\hss\box0\hss}
  \else
    \setbox1=\hbox to\wd0{\hss\box1\hss}
  \fi
  \mathop{\vcenter{
    \offinterlineskip\box0\box1}}}
\def\lesim{\m@thcombine<\sim}
\def\gesim{\m@thcombine>\sim}

\newcommand{\del}{\partial}
\newcommand{\colH}{{\cal H}_{\rm coll}}

\newcommand{\bg}{{\beta,\gamma}}

\begin{document}

\title{Quadrupole shape dynamics in view from a theory of large
amplitude collective motion}
\author{
  M. Matsuo\email{matsuo@nt.sc.niigata-u.ac.jp} \\
  \it Department of Physics, Faculty of Science, Niigata University, Niigata 950-2181, Japan  \\
  N. Hinohara \\
  \it Department of Physics and Astronomy, University of North Carolina,  Chapel Hill,\\ 
\it North Carolina 27599-3255, USA \\
  K. Sato, K. Matsuyanagi, T. Nakatsukasa,  \\
  \it RIKEN Nishina Center, Wako 351-0198, Japan \\
  K. Yoshida \\
  \it Graduate School of Science and Technology, Niigata University, Niigata 950-2181, Japan
}
\pacs{21.60.Ev,21.10.Re,21.60.Jz,27.30.+t,27.50.+e} 
\date{}
\maketitle

\begin{abstract}
  Low-lying quadrupole shape dynamics is a typical manifestation of large
  amplitude collective motion in finite nuclei. To describe the dynamics on
  a microscopic foundation, we have formulated a consistent scheme in 
  which the Bohr collective Hamiltonian
  for the five dimensional quadrupole shape variables is derived on the basis
  of the time-dependent Hartree-Fock-Bogoliubov theory. It enables us to 
  incorporates the Thouless-Valatin effect on the shape inertial functions,
   which has been neglected in previous microscopic Bohr Hamiltonian approaches.
  Quantitative successes are illustrated for the low-lying spectra in $^{68}$Se,
  $^{30-34}$Mg and  $^{58-64}$Cr, which display shape-coexistence, -mixing
  and -transitional behaviors.
   
\end{abstract}

\section{Introduction}

Recent nuclear structure studies develop significantly toward far from
the stability line as they are boosted by the RI beam facilities and the
advanced detector technologies. Low-lying quadrupole collectivity
is one of the highlights as characteristic spectra suggesting onset of
large deformation and coexistence of different shapes are often
observed in new regions of the experimental studies.  A
typical example is neutron-rich nuclei around $^{32}$Mg, in which
the lowering of the $2^+$ energy and the increase of the $B(E2)$
with increase of the neutron number indicate unusual onset of
quadrupole collectivity at the magic number $N=20$. Recent
identification of the second $0^+$ states at very low excitation
energy $\sim 1$ MeV \cite{Wimmer2010} poses further questions on the nature of this state,
for instance, whether it suggests coexistence of spherical
and prolately deformed states or not. Such a new region of
quadrupole collectivity is also found in neutron-rich Cr isotopes.

Theoretical description of excitation spectra associated with 
the shape coexistence and the shape transition is not a simple issue.
It is customary to consider the deformation energy surface by 
considering mean-field states with various shapes in
the $\beta-\gamma$ plane. However, the deformed states can rotate
and the deformation may evolve from one local minimum to 
others. One needs to describe this large amplitude dynamics.
Bohr's  five dimensional quadrupole coordinates
$a_{2\mu} (\mu=-2,-1,0,1,2)$, or equivalently the
$\beta-\gamma$ variables and the three Euler angles \cite{BohrVolII,Ring-Schuck} 
are suitable degrees of freedom for this purpose, but one then 
has to construct the collective Hamiltonian on the basis of the
nucleon many-body Hamiltonian.  This is a central problem
in this collective Hamiltonian approach, and several approaches
have been developed. 

The Bohr collective Hamiltonian consists of the collective potential 
$V(\beta,\gamma)$ and the kinetic energy term related to the
rotational and shape degrees of freedom, represented
by the three moments of inertia ${\cal J}_k(\beta,\gamma)$ with
$k=1,2,3$ being the principal axes of the deformation,
and the three shape inertial functions $D_{\beta\beta}(\beta,\gamma)$,
$D_{\gamma\gamma}(\beta,\gamma)$, and
$D_{\beta\gamma}(\beta,\gamma)$, which govern the kinetic energy
originating from the shape motion. In previous theories of microscopic 
Bohr Hamiltonian \cite{Pomorski1977,Dudek1980,Libert1999,Prochniak2004,
Niksik2009,Li2009,Prochniak2009}, however,  the Inglis-Belyaev 
cranking approximation is often adopted to evaluate the inertial functions, i.e.
 the velocity-dependent and time-odd mean fields induced by the 
 collective motion is neglected. This leads to an underestimate
of the inertial functions, and consequently to 
an overall stretching of 
 excitation spectra compared with the experimental observation \cite{Dudek1980,Prochniak2009}.
 If one includes the time-odd effect (the Thouless-Valatin effect) 
 on the rotational moments of inertia, the description of the yrast spectra is 
 improved \cite{Bertsch2007,Girod2009,Delaroche2010}, 
but leaving the problem in the yrare states such as 
 the second $0^+$ state. Clearly one should consider the Thouless-Valatin
 effect also on the shape inertial functions.

We have developed a microscopic theory of the quadrupole
collective dynamics which satisfies the above mentioned requirements \cite{Hinohara2010,Matsuyanagi2010,Matsuyanagi2013,Hinohara2011,Yoshida2011,Sato2012}.
It is based on a general theory of the large amplitude collective motion,
called the self-consistent collective coordinate (SCC) method \cite{Marumori1980,Matsuo2000}. 
The SCC method
starts with the time-dependent Hartree-Fock or time-dependent 
Hartree-Fock-Bogoliubov (TDHFB) theory that is powerful to describe 
many-body time-evolution of 
nuclei, including large amplitude motions such as a low-energy heavy-ion collision
 \cite{Flocard1978,Negele1982}. It 
provides a scheme to 
extract the collective submanifold
from the whole space of the TDHFB state vectors, and hence define
consistently collective coordinates and a collective Hamiltonian associated
with the collective submanifold. Applying
this method to the quadrupole shape dynamics, we succeeded in 
constructing the Bohr Hamiltonian from the microscopic many-body 
Hamiltonian \cite{Hinohara2010}. After briefly reviewing this theoretical scheme,
we illustrate how the theory solves the problem of
the stretched spectra, and works well for quantitative
description of the quadrupole shape dynamics.

\section{Adiabatic SCC method: theoretical backbone}

The time-evolution of the TDHFB state vector (a generalized
determinantal state) $ | {\phi(t)} \ket$
is given by the time-dependent variational principle
\begin{equation}
\delta\bra  \phi(t)| i\frac{\del}{\del t} -\hat{H} | \phi(t) \ket = 0.
\end{equation}
The SCC method \cite{Marumori1980} then assumes that the collective motion under consideration
corresponds to a subset of solutions of this equation, and that there exist
collective coordinates and momenta $(q_i,p_i) \ \ 
(i=1-M)$  that describe the collective motion. This means that
the time-evolution of the state vector is described via the time-evolution of
 $(q_i,p_i)$ which parametrize the TDHFB state vectors, i.e., 
$ | {\phi(t)} \ket =| \phi(q(t),p(t) )\ket$. The parametrized TDHFB state
vectors $\{ | \phi(q,p) \ket \}$ is called the collective submanifold. The 
time-evolution is governed by the collective Hamiltonian
$\colH(q,p) \equiv \bra  \phi(q,p) | \hat{H} | \phi(q,p) \ket$ and
the canonical equation of motion $\frac{d q_i}{dt} =\frac{\del \colH}{\del p_i},
\frac{d p_i}{dt} =-\frac{\del \colH}{\del q_i}$. Then, the time-dependent
variational principle Eq.(1) is transformed to
 the equation of the collective submanifold
\begin{equation}
\delta\bra  \phi(q,p)| \left( \frac{\del \colH}{\del p_i}\hat{P}^i + 
\frac{\del \colH}{\del q_i}\hat{Q}^i\right)  - 
\hat{H} | \phi(q,p) \ket = 0,
\end{equation}
where $\hat{P}^i$ and $\hat{Q}^i$ are displacement operators defined by
$ \hat{P}^i| \phi(q,p) \ket = i \frac{\del}{\del q_i}| \phi(q,p) \ket $
and $ \hat{Q}^i| \phi(q,p) \ket = - i \frac{\del}{\del p_i}| \phi(q,p) \ket $,
and they are one-body operators thanks to the Thouless theorem.
Equation (2) is self-contained.

To solve the equation, we introduce an expansion with respect to the
collective momenta $p$, assuming that the collective Hamiltonian
has a natural form
\begin{equation}
\colH(q,p)= \frac{1}{2}D(q)^{-1}_{ij}p_ip_j + V(q),
\end{equation}
consisting of the kinetic energy of the
collective motion with the inertial function $D(q)_{ij}$ and the collective potential
function $V(q)$. The state vector is expressed as 
$| \phi(q,p) \ket = \exp(ip_i\hat{Q}^i(q))|\phi(q)\ket$ using the Thouless theorem.
Expanding the equation of collective submanifold (2) with respect to the
powers of $p$, we obtain the following set of equations:
i) {\it the moving-frame HFB equation}
\begin{equation}
\delta\bra  \phi(q)| \hat{H}-\frac{\del V}{\del q_i}\hat{Q}^i(q) | \phi(q) \ket = 0,
\end{equation}
ii) {\it the moving-frame QRPA equation}
\begin{eqnarray}
&\delta\bra  \phi(q)|[ \hat{H}_M(q),\hat{Q}^i(q)] -D(q)^{-1}_{ij}\hat{P}^j(q)/i &\hspace{-5mm} | \phi(q) \ket = 0, \cr
&\delta\bra  \phi(q)|[ \hat{H}_M(q),\hat{P}^i(q)] -iC_{ij}(q)\hat{Q}^j(q) & \cr 
&\hspace{30mm} - \frac{\del V}{\del q_j}\Delta\hat{Q}^i_j(q) | \phi(q) \ket = 0,
\end{eqnarray}
together with the definitions of the collective potential
$V(q)=\bra  \phi(q)| \hat{H} | \phi(q) \ket$, the collective
inertial function $D(q)^{-1}_{ij}=-\bra  \phi(q)| [[\hat{H},\hat{Q}^i(q)],\hat{Q}^j(q)] | \phi(q) \ket$,
and the moving-frame Hamiltonian 
$\hat{H}_M(q)= \hat{H}-\frac{\del V}{\del q_i}\hat{Q}^i(q)$.
Apart from the curvature term $\Delta\hat{Q}(q)$, Eq.(5) is similar to
the 
QRPA equation determining a normal mode.
  Note that $\hat{Q}^i(q)$ also plays a role of
 the constraining operator in the moving-frame HFB equation, and
hence the two equations are coupled. These are basic equations
of the adiabatic SCC (ASCC) method \cite{Matsuo2000}.
 The ASCC method is a natural extension of the
 RPA theory, with which one can extract normal modes of small amplitude
 oscillation.

\section{CHFB plus local QRPA approach: a practical implementation}
 
Our scheme to derive the Bohr collective Hamiltonian is a practical and approximate
implementation of the ASCC method \cite{Hinohara2010}. We first perform a standard 
 constrained  Hatree-Fock-Bogoliubov (CHFB) calculation to obtain 
mean-field states $|\phi_{\rm CHFB} (\beta,\gamma)\ket$ with various quadrupole 
deformations in the $\beta-\gamma$ plane. We assume that the 
collective submanifold states $|\phi(q)\ket$ have a one-to-one mapping to the 
CHFB states $|\phi_{\rm CHFB} (\beta,\gamma)\ket$ by a coordinate transformation.
The collective potential is then the
deformation energy of the CHFB states
$V(\beta,\gamma)=  
\bra  \phi_{\rm CHFB}(\beta,\gamma)| \hat{H} | \phi_{\rm CHFB}(\beta,\gamma) \ket$.
For the moving-frame QRPA equation (5), we neglect
the curvature term
$\Delta \hat{Q}$ and replacing $\hat{H}_M(q)$ with the constrained Hamiltonian
$\hat{H}_{\rm CHFB}$, and we solve
\begin{eqnarray}
\delta\bra  \phi_{\rm CHFB}(\beta,\gamma)|[ \hat{H}_{\rm CHFB},\hat{Q}^i] -D^{-1}_i\hat{P}^i/i | \phi_{\rm CHFB}(\beta,\gamma) \ket = 0, \cr
\delta\bra  \phi_{\rm CHFB}(\beta,\gamma)|[ \hat{H}_{\rm CHFB},\hat{P}^i/i] -C_i\hat{Q}^i
| \phi_{\rm CHFB}(\beta,\gamma) \ket = 0.
\end{eqnarray}
We call this local QRPA (LQRPA) equations. They determine the displacement
operators $\hat{Q}^i$ and $\hat{P}^i$, and hence the  vibrational 
collective coordinates $(q_i, p_i)$, which are orthonormal  
locally at each $(\beta,\gamma)$ point. 
($C_{ij}$ and $D^{-1}_{ij}$ are diagonal then, and for simplicity we often choose the scale
satisfying $D_i=1$.)
Among various QRPA solutions
we select two which are most effective to change the quadrupole deformations.
The collective coordinates of the normal modes
$(q_i, p_i)$ $(i=1,2)$ can be related to
 the Bohr  quadrupole shape variables 
$a_{20} = c\bra \hat{Q}_{20}\ket ,a_{22}=c\bra \hat{Q}_{22}\ket$
via
\begin{equation}
\frac{\del a_{2m}}{\del q_i}=c\bra  \phi_{\rm CHFB}(\beta,\gamma)| 
[\hat{Q}_{2m}, \hat{P}^{i}/i] | \phi_{\rm CHFB}(\beta,\gamma) \ket.
\end{equation}
The vibrational kinetic energy 
$T_{\rm vib}=(p_1^2+p_2^2)/2=(\dot{q}_1^2+\dot{q}_2^2)/2$ is then expressed 
in the Bohr  coordinates as
\begin{eqnarray}
T_{\rm vib}=\frac{1}{2}\sum_{mm'=0,2}M_{mm'}\dot{a}_{2m}\dot{a}_{2m'} 
\hspace{30mm} \nonumber\\
\ \ \ =\frac{1}{2}D_{\beta\beta}(\beta,\gamma)\dot{\beta}\dot{\beta}
+D_{\beta\gamma}(\beta,\gamma)\dot{\beta}\dot{\gamma}
+\frac{1}{2}D_{\gamma\gamma}(\beta,\gamma)\dot{\gamma}\dot{\gamma},
\end{eqnarray}
with
\begin{equation}
M_{mm'}=\sum_{i=1,2}\frac{\del q^i}{\del a_{2m}}\frac{\del q^i}{\del a_{2m'}},
\end{equation}
and the inertial functions $D_{\beta\beta},D_{\beta\gamma}$ and 
$D_{\gamma\gamma}$ are derived from $M_{mm'}$ by a straight
variable transformation from $(a_{20},a_{22})$ to $(\beta,\gamma)$.
We calculate the rotational moments of inertia also by using the local
QRPA equation
\begin{equation}
\delta\bra  \phi_{\rm CHFB}(\beta,\gamma)|
[ \hat{H}_{\rm CHFB},\hat{\Theta}_k] -{\cal J}_k^{-1}\hat{I}_k/i 
| \phi_{\rm CHFB}(\beta,\gamma) \ket = 0,
\end{equation}
$(k=1,2,3)$ for the rotation around the three principal axes, and obtain
the rotational kinetic energy
$
T_{\rm rot}=\frac{1}{2}\sum_k {\cal J}_k(\beta,\gamma) \omega_k^2.
$

We do not neglect any residual interactions in solving the
LQRPA equations, i.e. we take into account all the induced fields,
including the time-odd ones, associated with the collective rotations
and the $\beta-\gamma$ shape motions. The Thouless-Valatin
effects are thus taken into account for all the inertial functions
${\cal J}_x,{\cal J}_y,{\cal J}_z,D_{\beta\beta},D_{\beta\gamma}$ and 
$D_{\gamma\gamma}$. If the residual interactions were neglected
 in solving Eqs.(6) and (10),
the approximation would lead  to 
the Inglis-Belyaev inertial functions. Although the Thouless-Valatin effect 
on the rotational moments of inertia is widely known, and is taken
into account in a recent approach \cite{Bertsch2007,Girod2009,Delaroche2010},
the systematic inclusion of the Thouless-Valatin effect on
the vibrational inertia is achieved in our approach for the first time.

\section{Quadrupole dynamics with the LQRPA inertial functions}

In applications of the CHFB + LQRPA, we have to solve
the QRPA equations for all the CHFB states 
with various deformations.
To construct the CHFB states and the QRPA solutions, we employ
the pairing plus quadrupole model. Two major shells are adopted
as a model space for each of neutrons and protons.  The
single-particle energies, the force parameters of the monopole
pairing interaction and the quadrupole-quadrupole interaction
are adjusted to reproduce the results of the Skyrme-HFB calculation.
The quadrupole pairing interaction is also taken into account  since
it is known to bring about the Thouless-Valatin effect \cite{Sakamoto1990}.
We use the selfconsistent strength 
of the quadrupole pairing, with which the Galileian invariance of
the pairing interaction is recovered \cite{Sakamoto1990}. We first obtain the CHFB states 
$ | \phi_{\rm CHFB}(\beta,\gamma) \ket$ for $60 \times 60$ mesh points,
discretized both for $\beta=0 - \beta_{\rm max}$ (typically $\beta_{\rm max}=0.6$)
and for $\gamma=0-\pi/3$. 
The QRPA equations are then solved  the same number of times. 

To obtain the excitation spectra of the quadrupole dynamics, we requantize
the Bohr Hamiltonian using the standard Pauli prescription:
\begin{eqnarray}
 \{\hat{T}_{\rm vib} + \hat{T}_{\rm rot} + V \}
 \Psi_{\alpha IM}(\bg,\Omega) = E_{\alpha I} \Psi_{\alpha IM}(\bg,\Omega).
\label{eq:Schroedinger}
\end{eqnarray}
Expressing the 
collective wave function
$
\Psi_{\alpha IM}(\beta,\gamma,\Omega)=\sum_{K}\Phi_{\alpha IK}(\beta,\gamma)
\bra \Omega|IMK\ket
$
directly on the  $60 \times 60$ mesh points in the $\beta-\gamma$ plane, we
diagonalize a Hamiltonian matrix in the mesh representation.
 $\Omega$ is the Euler angle, and $\bra \Omega|IMK\ket$ is the rotational function.

\subsection{Spectra in $^{68}$Se and the Thouless-Valatin effect}

The first application of the CHFB + LQRPA approach was performed for
$^{68}$Se \cite{Hinohara2010}, in which the experimental spectra suggest possible 
coexistence of
oblate and prolate shapes. The calculated CHFB potential energy surface
indeed shows two minima at prolate shape $\beta\sim 0.3, \gamma=0$
  and at oblate shape $\beta\sim 0.3, \gamma=\pi/3$, but the potential energy
  surface is soft with respect to the $\gamma$ direction with the energy difference 
of several hundreds keV and a few hundreds keV barrier.

The Thouless-Valatin effect is examined by comparing the LQRPA
inertia, for instance, $D_{\beta\beta}$, with the same quantity
$D_{\beta\beta}^{(IB)}$ evaluated
in the Inglis-Belyaev cranking approximation. We have found 
that the ratio $D_{\beta\beta}/D_{\beta\beta}^{(IB)}$ is typically 1.3-1.5 in a 
large part of the $\beta-\gamma$ plane, where the deformation energy is
not large. At larger deformation $\beta \gesim 0.4$ it takes
values around 2. Concerning the rotational  moment of inertia,  
the ratio ${\cal J}_1 /{\cal J}_1^{(IB)}$ 
takes a value around $1.2-1.5$ at small and modest deformation, 
but it decreases to $\sim 1.1$ at
large deformations. The deformation dependence  is different between
the LQRPA inertia and the IB inertia. Also the six inertial functions have
different deformation dependences.

The Thouless-Valatin effects on the inertial functions have impact on the
excitation spectra. When we perform a calculation using the Inglis-Belyaev
cranking inertia, the excitation energies of the yrast states
are $E(2_1^+), E(4_1^+),  E(6_1^+)=0.991, 2.310, 3.891$ MeV, which are
systematically larger than the corresponding experimental values
$0.854, 1.942, 3.304$ MeV by about 20\%.
This deficiency is improved in the CHFB+LQRPA calculation,
producing $E(2_1^+), E(4_1^+),  E(6_1^+)=0.810, 1.951, 3.348$ MeV
in much better agreement with the experiment.  The improvement
is achieved also for the yrare states; the CHFB+LQRPA description
gives the second $2^+$ state at $E(2_2^+)=1.536$ MeV consistent with
the experimental value $E(2_2^+)=1.593$ MeV, and much better than the
calculation $E(2_2^+)=1.883$ MeV using the Inglis-Belyaev inertia.

\subsection{Collective wave function and the shape fluctuation in
neutron-rich isotopes near $^{32}$Mg}

\begin{figure}
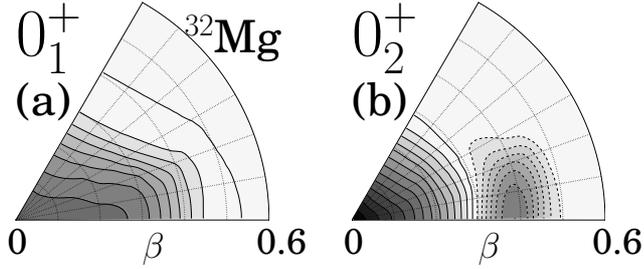

\begin{center}
\includegraphics[width=4.3cm]{32Mg-I0K0k0.eps}
\includegraphics[width=4.3cm]{32Mg-I0K0k1.eps}
\end{center}
\caption{
Vibrational wave functions 
$\Phi_{\alpha IK}(\beta,\gamma)$
of the $0_1^+$ and $0_2^+$ states in $^{32}$Mg.
Dotted curves are contours for negative values.
}
\end{figure}

The nature of the quadrupole collectivity can be examined by analyzing
the collective wave function Eq.(11) and its distribution in the $\beta-\gamma$
plane. An interesting example is neutron-rich Mg isotopes around $A=32$ \cite{Hinohara2011}.
The region of these isotopes are known as an island of inversion since 
a large quadrupole collectivity is observed in spite of the neutron magic
number $N=20$. 
Indeed a steep change of the energy ratio, $E(4_1^+)/E(2_1^+)=$ 2.23, 2.64 and 3.21 in experiment
and 2.37, 2.82 and 3.26 in theory for
$A=30,32$ and 34, indicates a dramatic evolution from vibrational to rotational
behaviors. 
The yrare $0_2^+$ state
is observed also in $^{30,32}$Mg, at very low energies close to $E(2_1^+)$.
The calculation
 reproduces well the energies of the yrast $2_1^+$ and $4_1^+$ states 
and the yrare $0_2^+$ state; they are 0.744, 2.099 and 0.986 MeV in $^{32}$Mg, compared with
the experimental values, 0.885, 2.322 and  1.058 MeV, respectively.

The obtained collective wave function for the ground $0^+_1$ state 
exhibits a clear shape transition from a spherical state in $^{30}$Mg
(a distribution concentrated around $\beta\sim 0$),
to a well deformed state in $^{34}$Mg (a distribution around $\beta\sim 0.35, \gamma\sim0$).
In $^{32}$Mg, the wave function of the ground state 
indicates a significant fluctuation in the quadrupole shape as it
is widely spread over $\beta = 0 \sim 0.4$ along
the $\gamma=0$ line (Fig.1(a)).
The nature of the $0_2^+$ states is also interesting. Experimentally,
the $0_2^+$ state in $^{32}$Mg is populated strongly by the two-neutron
transfer $(t,p)$ reaction from $^{30}$Mg  \cite{Wimmer2010}.
It has been argued that the $0_1^+$ and $0_2^+$ states in $^{30,32}$Mg are coexisting
states, which are either spherical or deformed, and interchanging each other across $N=20$.
Our theoretical picture is different. The wave function of
the  $0^+_2$ state  is  spread over a large
interval $\beta=0 \sim 0.45$, but it has a node around $\beta \sim 0.3$ (Fig.1(b)). 
It indicates a significant shape fluctuation likewise in the ground state, and it
contains to some extent a feature of the $\beta$-vibration.
(The character of the $\beta$-vibration in the $0_2^+$ state develops well in $^{34}$Mg.)
This is far from the simple shape coexistence picture. 

\subsection{Shape transition in neutron-rich Cr isotopes}

\begin{figure}
\begin{center}
\includegraphics[width=8cm]{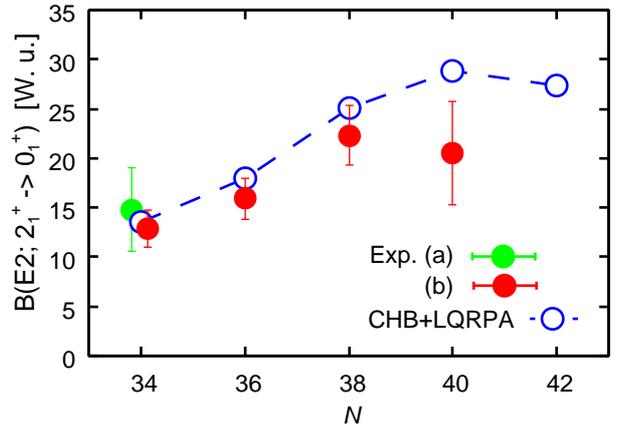}
\end{center}
\caption{
Comparison of theoretical and experimental $B(E2; 2_1^+ \rightarrow 0_1^+)$ in
the Weisskopf unit for the Cr isotopes. The experimental values are taken from 
Refs. \cite{Burger2005-Cr,Baugher2012-Cr,Crawford2013-Cr} while the 
CHFB+LQRPA value is from Ref. \cite{Sato2012}.
}
\end{figure}

Neutron-rich Cr isotopes with $A\gesim 60$ are another example of
the new regions of deformation, which is suggested in recent experiments.
CHFB calculations using the Skyrme functional (SkM*) for axial deformations
indicate that a softening of the deformation potential around $N=34$ 
develops further with increasing the neutron number, and producing a
deformed but shallow minimum at $\beta \sim 0.3$ for $N \ge 38$  \cite{Oba2008,Yoshida2011}.
The calculated excitation energy of the $2_1^+$ state decreases
gradually with increasing the neutron number \cite{Sato2012}. The energy ratio 
$E(4_1^+)/E(2_1^+)$ gradually increases from 2.12 at $N=34$
to 2.68 at $N=40$, and  similarly in $B(E2; 2_1^+ \rightarrow 0_1^+)$ 
as shown in Fig. 2. The isotopic trend of $B(E2)$ values is
 in good agreement with the data.
The absolute values of the excitation energy are also well described;
$E(2_1^+), E(4_1^+)=0.502, 1.228 $ MeV in theory v.s.
0.446, 1.180 MeV in experiment for $^{62}$Cr. 

From the above quantities one clearly sees the
 gradual development of quadrupole collectivity.
However the yrast spectra display a
transitional behavior between the spherical vibrator and the deformed rotor
 even in the isotopes with 
$N=40,42$ having the largest collectivity.
We can learn more from the 
collective wave functions. Indeed, we found that  the wave functions
of the yrast states $0_1^+,2_1^+, 4_1^+$, especially at lower spin members,  
spread largely from the prolate
shape $\beta\sim 0.5, \gamma\sim 0$ toward the oblate shape. 
More significant influence of the $\gamma$ degrees of freedom is predicted
in the yrare $K^\pi=0^+$ and $2^+$ bands in $^{64}$Cr.

\section{Conclusions and perspectives}

The CHFB + LQRPA approach provides us with a 
scheme to construct the Bohr collective Hamiltonian for the large
amplitude quadrupole shape motion on the
basis of the microscopic many-body mean-field dynamics.  
A great advantage of this approach is that the effect of the
velocity-dependent (time-odd) induced field on the inertia of the
collective motion, i.e. the Thouless-Valatin effect, is taken
into account not only for the rotational moments of inertia 
but also for the vibrational inertia with respect to the $\beta-\gamma$ 
shape motion. This is achieved in solving the  local QRPA equation at each deformation.
It seems to solve the problem of the 
Inglis-Belyaev inertia that often produces stretched spectra.

Further developments of the CHFB+LQRPA approach are anticipated.
We would like to perform calculations on the basis of the 
selfconsistent Hartree-Fock-Bogoliubov models using the modern energy density functional,
for instance, the Skyrme functional. Constrained HFB calculations in two dimensional
deformation spaces such as the $\beta-\gamma$ plane are numerically intensive, but
manageable on reasonable computers.  A difficulty lies in the local QRPA part. The
QRPA calculation using the realistic density functional requires a large single-particle
space to guarantee the selfconsistency, allowing, at present, calculations assuming 
axially symmetric deformations.  The QRPA calculation
with non-axial deformations will be very tough in numerics, and we have to
perform the calculations for all the deformations. This is a challenge, but 
it will become feasible in near future thanks to the rapid development of computer powers. 
Application to the spontaneous fission will be within a scope then.

\begin{ack}
This work is supported by 
KAKENHI (Nos. 21340073, 23540294, 23740223 and 25287065).
\end{ack}


\end{document}